\documentstyle[prl,tighten,multicol,aps,epsfig]{revtex}

\begin{document}
\draft
\preprint{}

\title{
Quantum Phase Transition in Extended Attractive Hubbard Model
}
\author{S. Saito,$^1$ H. Yoshimoto,$^1$ Y. Y. Suzuki,$^2$ and S. Kurihara$^1$}
\address{
$^1$Department of Physics, Waseda University, 3-Okubo, Shinjuku-ku, Tokyo 169-8555, Japan \\
$^2$NTT Basic Research Laboratories, Atsugi 243-0198, Japan
}

\date{\today}
\maketitle
\begin{abstract}
We have made a variational analysis on an evolution of superconductivity from weak to strong coupling regime.
In contrast to a crossover without thermodynamic anomaly found in a dilute system, we show the existence of a quantum phase transition near half filling.
The transition is driven by charge density waves instabilities.
We have found that superconductivity and charge density waves coexist in the presence of a weak intersite repulsion.
The ground state phase diagram is determined and the quantum phase transition is an attractive version of the Mott transition.  
\end{abstract}



\begin{multicols}{2}
\narrowtext

The Landau Fermi liquid theory~\cite{Landau57} explains thermodynamic properties and responses to external fields of interacting fermion systems, and predicts the existence of collective excitations.
It is based on the assumption of {\it adiabatic continuity} from the noninteracting system.
However, this assumption is not always satisfied in strongly interacting electron systems.
On increasing repulsion, a metallic Fermi liquid (FL) phase breaks down, and the system becomes an antiferromagnetic (AF) insulator~\cite{Mott90}.  
This is the Mott transition --- a kind of {\it quantum phase transition}.
We investigate whether a corresponding transition exists in a superconducting (SC) phase on increasing {\it attraction} instead of repulsion.

The Bardeen Cooper Schrieffer (BCS) theory has been the basis for our understanding of superconductivity and is known to be valid in the weak coupling regime~\cite{Bardeen57}.
We think that the BCS theory in the strong coupling regime should be critically examined in relation with the small coherence length superconductors discovered in the last decades~\cite{Micnas90}.
In the strong coupling case, tightly bound pairs of electrons are expected to behave like Bose particles called bipolarons, and Bose-Einstein condensation (BEC) should take place.
According to the work of Nozi$\grave{{\rm e}}$res and Schmitt-Rink~\cite{Nozieres85} the evolution with increasing attraction is a {\it smooth crossover}.
This result is widely accepted, at least in qualitative sense.
The smooth evolution scenario is justified in a dilute system, but must break down near half filling $(n=1)$ where a charge density waves (CDW) phase is expected.
We reexamine the evolution taking the possible occurrence of CDW phase into account, and find a quantum phase transition between SC and CDW.
Moreover, we find that these different kinds of orders compete, and {\it coexist} under certain conditions.
A coexistence of SC and CDW has been recently observed in a stripe phase of high $T_{\rm c}$ superconductors~\cite{Tranquada95,Suzuki98,Hunt99}.

We employ an infinite dimensional $(d=\infty)$ extended attractive Hubbard model on an AB-{\it bipartite} hypercubic lattice~\cite{Metzner89,Muller-Hartmann89,Georges96,Freericks94,Ohkawa92,Taraphder95} to simplify the problem, without losing essential feature of quantum fluctuations.
\begin{eqnarray}
H&=&-\frac{t^{*}}{\sqrt{2d}}\sum_{\langle ij \rangle \sigma}(c_{i\sigma}^{\dagger}c_{j\sigma} + {\rm H. C.})
- U \sum_{i}n_{i\uparrow}n_{i\downarrow} \nonumber \\
&&{}+ \frac{V^{*}}{d}\sum_{\langle ij \rangle \sigma \sigma^{\prime}}n_{i\sigma}n_{j\sigma^{\prime}}. \label{tUV}
\end{eqnarray}
The last term represents a nearest neighbor repulsion which plays an important role to stabilizing CDW.

To improve the BCS theory in a controlled way, we rely on a variational method as a guiding principle for the ground state.  
We consider the following Gutzwiller mean-field variational wave function (GMF)
\begin{equation}
	|{\rm GMF} \rangle = \prod_{i} g_{i}^{n_{i\uparrow}n_{i\downarrow}-\mu_{i\uparrow}n_{i\uparrow}-\mu_{i\downarrow}n_{i\downarrow}+\eta_{i}} g_{i}^{-\nu_{i} c_{i\downarrow}c_{i\uparrow}} |\rm{MF} \rangle \label{WF},
\end{equation}
where $|\rm{MF} \rangle$ is a mean-field (MF) wave function.
This is an extension of the Gebhard wave function~\cite{Gebhard90} to include SC order with $|\rm{MF} \rangle$.
Parameters $\mu_{i\sigma}$, $\eta_{i}$, and $\nu_{i}$ are determined so as to eliminate any self-energy diagrams in $d=\infty$.
The Gutzwiller projection incorporates local electronic correlations, and $g_{i}>1$ in an attractive system.
We have found significant deviations from the BCS theory with increasing $U$, and $g_{i}\rightarrow 2$ as $U\rightarrow \infty$~\cite{Suzuki99}.

\begin{figure}[thb]
\begin{center}
\epsfig{file=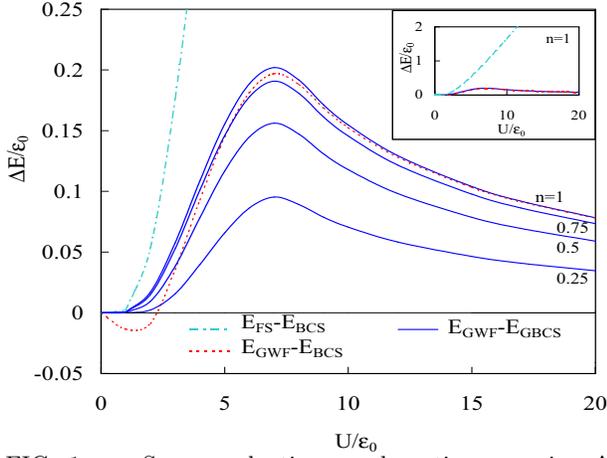,width=8cm,height=6cm}
\caption{
Superconducting condensation energies $\Delta E $ $\equiv E_{\rm FL}-E_{\rm SC}$ of the attractive Hubbard model at $n=1$, $0.75$, $0.5$, $0.25$, compared with several differences at $n=1$.
The inset illustrates the large scale behavior.
} 
\label{DE_SC}
\end{center}
\end{figure}

Before discussing SC we comment on the normal state.
We substitute the noninteracting Fermi sea (FS) for $|\rm{MF} \rangle$, and the resulting $|{\rm GMF} \rangle$ is the Gutzwiller wave function $|\rm{GWF} \rangle$.
On increasing $U$, the discontinuity $q$ at the Fermi surface of the momentum distribution function diminishes.
This means that the quasi-particles effective mass increases.
At $U=U_{\rm BR}\equiv8\epsilon_{0}$ and arbitrary filling $n$, with $\epsilon_{0}=\sqrt{2/\pi}t^{*}$ the average band width, a quantum phase transition arises.
We obtain $q\rightarrow 0$ for $U>U_{\rm BR}$.
The electrons completely localize forming bipolarons, where the spin susceptibility $\chi_{\rm s} \rightarrow 0$ and a spin gap opens up.
This quantum phase transition corresponds to the Brinkman-Rice metal-insulator transition in the repulsive case~\cite{Brinkman70}.
It does not actually occur if we take SC into account.
However, it implies that a crossover from FL to a Bose liquid takes place at $U \sim U_{\rm BR}$.
This interpretation is consistent with the peak positions of superconducting condensation energies, as shown in Fig. \ref{DE_SC}.

For SC we substitute in the BCS wave function $|{\rm BCS} \rangle$ for $|{\rm MF} \rangle$ in Eq. (\ref{WF}), and we call it the Gutzwiller BCS wave function $|{\rm GBCS} \rangle$.
We obtain an optimized variational energy with the $s$-wave symmetry~\cite{Suzuki99}.
We show, in Fig. \ref{DE_SC}, SC condensation energies $\Delta E\equiv E_{\rm FL}-E_{\rm SC}$ of the attractive Hubbard model $(V^{*}=0)$.
$\Delta E$ increases following the BCS formula $\sim t^{*} \exp(-\alpha t^{*}/U)$ in the weak coupling regime, where $\alpha$ is of the order unity, decreases proportional to $t^{* \ 2}/U$ in the strong coupling regime, and has a peak near $U\sim U_{\rm BR}$ independent of $n$.
Simple $|{\rm BCS} \rangle$ and $|{\rm FS} \rangle$ cannot reproduce this behavior, and we must take Fermi liquid effects into account for both FL and SC.
Therefore, the work of Nozi$\grave{{\rm e}}$res and Schmitt-Rink~\cite{Nozieres85} is insufficient due to a lack of Fermi liquid effects.
Note that this by no means suggests a break-down of the BCS theory.
In the original BCS theory, FL is considered as the normal state, and shows that FL is unstable towards SC in the presence of the infinitesimally weak attraction between {\it quasi-particles}~\cite{Bardeen57}.
Then, operators in the BCS-reduced-Hamiltonian, should be considered as dressed {\it quasi-particles}.
Instead, we start from the attractive Hubbard model represented by {\it bare} electrons, and incorporate Fermi liquid effects by the Gutzwiller projection.
We have shown that an infinitesimally weak attraction actually leads to SC by $\rm | GBCS\rangle$.
$\Delta E$ is a reminiscent of an expected SC transition temperature $T_{\rm c}$ in a crossover from BCS to BEC~\cite{Micnas90,Nozieres85,Freericks94,Keller99}.

\begin{figure}[htb]
\begin{center}
\epsfig{file=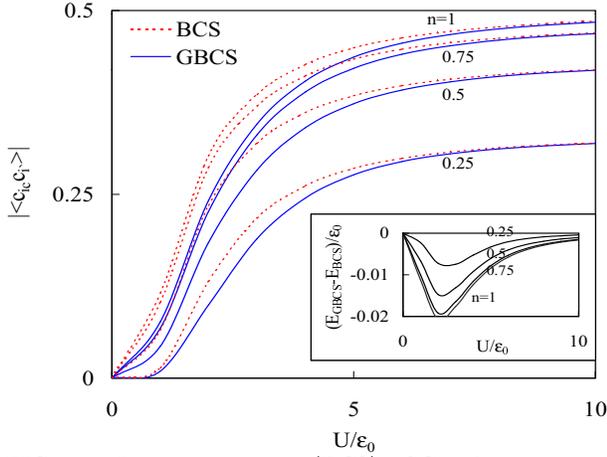,width=8cm,height=6cm}
\caption{
Improvements on $\rm | BCS\rangle$.  
SC order parameters calculated with $\rm | GBCS\rangle$ are shown at $n=1, 0.75, 0.5, 0.25$, combined with the BCS results (the dotted line).
The inset is the energy difference.
} 
\label{BCSvsGBCS}
\end{center}
\end{figure}

We show in Fig. \ref{BCSvsGBCS}, the SC order parameter $|\langle c_{i\downarrow}c_{i\uparrow}\rangle|$ and the energy difference between $\rm | BCS\rangle$ and $\rm | GBCS\rangle$.
Deviations are most prominent in the intermediate coupling regime.
SC order parameters are overestimated by $\rm | BCS\rangle$ compared with $\rm | GBCS\rangle$.
In $\rm | BCS\rangle$, variational energies are lowered only by the presence of an SC order parameter, whereas in $\rm | GBCS\rangle$, we can further lower variational energies by Fermi liquid effects.
The agreement in a weak coupling limit is expected, but the fact that they also agree in a strong coupling limit, is a result of $d=\infty$ limit.
In a strong coupling limit $U\rightarrow \infty$, the attractive Hubbard model maps onto the spin model~\cite{Micnas90,Matsubara56}, i.e. the Heisenberg model with an external magnetic field.
However, in $d=\infty$, nearest neighbor spin-spin interaction reduces to only Hartree contributions~\cite{Muller-Hartmann89}, and a MF treatment becomes exact.
We have checked correct MF limits, $\langle n_{i\uparrow}n_{i\downarrow}\rangle \rightarrow n/2$ and $|\langle c_{i\downarrow}c_{i\uparrow}\rangle | \rightarrow \sqrt{n(2-n)}/2$, as $U\rightarrow \infty$.
In finite dimensions $d<\infty$, the simple spin waves theory ($1/S$ expansion) is sufficient to give lower order parameters compared with a MF theory, and we expect that deviations persist finite as $U\rightarrow \infty$.
Although $\rm | GBCS\rangle$ shows a significant deviation from $\rm | BCS\rangle$, as far as a pure SC is concerned, the evolution turns out to be continuous, which is consistent with common wisdom.

Now, we explore a {\it discontinuous} evolution.
SC and CDW are characterized by the diagonal and off-diagonal long range order in the Nambu density matrix
\begin{eqnarray}
\langle \psi_{i}^{\dagger } \psi_{j} \rangle 
 &\equiv& 
\pmatrix{
\langle c_{i\uparrow}^{\dagger}c_{j\uparrow} \rangle & \langle c_{i\uparrow}^{\dagger}c_{j\downarrow}^{\dagger} \rangle \cr
\langle c_{i\downarrow}c_{j\uparrow} \rangle & \langle c_{i\downarrow}c_{j\downarrow}^{\dagger} \rangle 
}, \label{Nambu}
\end{eqnarray}
and we investigate a competition of these orders.
In order to take a coexistence phase (SC+CDW) into account, we put 
\begin{equation}
|{\rm SC}+{\rm CDW} \rangle=
\prod_{{\bf k}} 
	[
	u^{\rm s}_{\bf k}+v^{\rm s}_{\bf k}
				\alpha_{{\bf k}\uparrow}^{\dagger}\alpha_{{\bf -k}\downarrow}^{\dagger}
	]|0 \rangle	\label{MF_SC+CDW}
\end{equation}
as $|{\rm MF} \rangle$, where $\alpha_{{\bf k}\sigma}$ denotes an annihilation operator for CDW bogolons defined by
\begin{eqnarray}
\pmatrix{
	\alpha_{{\bf k}\sigma} \cr
	\alpha_{{\bf k+Q}\sigma}
}=
\pmatrix{
	u^{\rm c}_{\bf k}	&	v^{\rm c}_{\bf k} \cr
	-v^{\rm c}_{\bf k}	&	u^{\rm c}_{\bf k}
}
\pmatrix{
	c_{{\bf k}\sigma} \cr
	c_{{\bf k+Q}\sigma}
},	\label{CDWbogolon}
\end{eqnarray}
and ${\bf Q}=(\pi, \pi, \cdots)$ is a commensurate CDW wave vector.
Our wave function $|{\rm GMF} \rangle$ comes from a natural unification of the spin-bag approach~\cite{Schrieffer89} and the resonating-valence-bond idea~\cite{Anderson87} for the attractive system.
Optimal functional forms of variational parameters $u$ and $v$ are analytically determined beyond a MF theory, and further minimizations are performed numerically~\cite{Suzuki99}.

In the absence of $V^{*}$, CDW and SC are degenerate at $n=1$, and minimum variational energies do not change for a coexistence phase.
This result was obtained in the MF theory~\cite{Micnas90} and is not altered by the Gutzwiller projection.
This reflects a symmetry~\cite{Zhang90} of the attractive Hubbard model.
However, such a macroscopic degeneracy is lifted under any weak perturbation.  
In fact, for $V^{*} > 0$ we obtain a pure CDW ground state at $n=1$, and we find a coexistence phase at $n\neq 1$.

\begin{figure}[thb]
\begin{center}
\epsfig{file=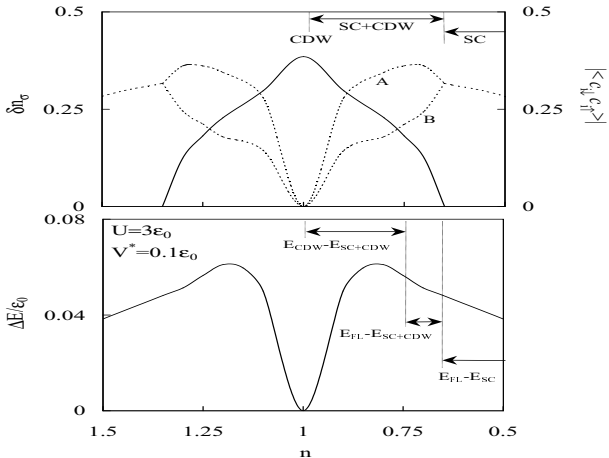,width=8cm,height=6cm}
\caption{
Coexistence of SC and CDW at $U=3\epsilon_{0}$ and $V^{*}=0.1\epsilon_{0}$.
A coexistence phase is denoted by SC+CDW, and pure phases are denoted by SC or CDW.
The upper Fig. shows order parameters of CDW $\delta n_{\sigma}=(n_{{\rm A}\sigma}-n_{{\rm B}\sigma})/2$ (the solid line) and SC $|c_{i\downarrow}c_{i\uparrow}|$ (the dotted line, A and B refer to the sublattice).
The lower Fig. shows superconducting condensation energies.
} 
\label{Order}
\end{center}
\end{figure}

We show, in Fig. \ref{Order}, order parameters and variational energies as functions of $n$.
CDW suppresses SC but these orders can coexist near $n=1$.
Doped holes in CDW are expected to behave like Bose particles in a strong coupling regime.
As far as concentrations of holes are dilute, they can not completely destroy CDW.
Moreover, a dilute Bose gas exhibits BEC at zero temperature, and as a result the coexistence is realized.
Further reduction of $n$ induces a quantum melting of CDW, and we find a pure SC.
The quantum phase transition is of the second order.

A coexistence of SC and {\it charge} stripe orders was observed by NQR measurements in cuprates~\cite{Hunt99}.
Although dimensionalities and CDW structures are quite different from the experiments, our results qualitatively agree with observed behaviors of order parameters.
The coexistence of $d$-wave SC and AF has also been found by the variational Monte Carlo method in the two dimensional $t$-$J$ model~\cite{Chen90}.
Neutron scattering experiments of cuprates revealed that three orders (SC, CDW, and AF) coexist near $1/8$ hole doping~\cite{Tranquada95}.
Therefore, our results are complementary to previous theoretical results and are supported by experimental observations.
A coexistence phase is also observed in bismuthate superconductors by infrared measurements~\cite{Blanton93}.
However, in bismuthates, SC is realized after several tens of percent of doping~\cite{Taraphder95,Blanton93}.
To understand these experiments, we must take into account a finite dimensionality, incommensurate CDW, localization effects due to impurities, and so on.

Another possibility for the ground state is a phase separation~\cite{Batrouni00} between electrons rich and poor regions.
In fact, the inverse of the charge susceptibility $\chi_{c}^{-1}=\partial  \mu /\partial n \sim 0$ shows that the system is a proximity of it.
However, unlike neutral bosons, charged electrons are hard to exhibit a macroscopic phase separation due to the long range Coulomb interaction.

\begin{figure}[thb]
\begin{center}
\epsfig{file=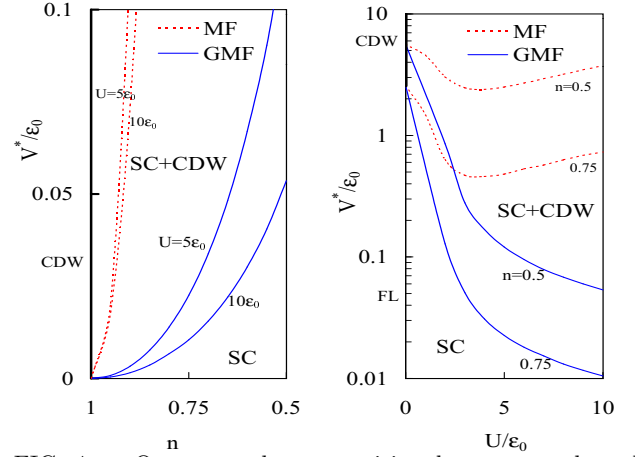,width=8cm,height=6cm}
\caption{
Quantum phase transition between weak and strong coupling superconductivity.
The solid line is the phase boundary of a pure SC and a coexistence phase (SC+CDW).
The dotted line is the MF result.
The left Fig. is obtained at $U=5\epsilon_{0}$, $10\epsilon_{0}$.
At $n=1$, a pure CDW is stable for $V^{*}>0$.
The right Fig. is obtained at $n=0.75$, $0.5$.
Normal states are stable at $U=0$, and there exists a transition between a homogeneous FL and a pure CDW.
} 
\label{Vc(n)}
\end{center}
\end{figure}

We show in Fig. \ref{Vc(n)}, phase diagrams at fixed values of $U$ or $n$.
Clearly, CDW are more stable near $n=1$.
The MF theory underestimates CDW instability.
A pure CDW is stable at $n=1$ and $V^{*}>0$.
We find a quantum phase transition between a pure SC and a coexistence phase by varying $U$ and/or $V^{*}$.
$\rm | MF\rangle$ gives a reentrant behavior of the phase boundary.
In contrast, $\rm | GMF\rangle$ yields smaller value of $V^{*}$ for CDW with increasing $U$.
Our results can be understood by a simple picture: on increasing $U$ coherence length of Cooper pairs becomes smaller, leading to a smaller value of $V^{*}$ for CDW due to larger susceptibility of different pairs.
This behavior is also expected from an effective Hamiltonian in the strong coupling regime, the anisotropic Heisenberg model~\cite{Micnas90}: SC and CDW orders correspond to ferromagnetism in the $xy$ plain and AF along the $z$ direction respectively, and the effective exchange interactions are $J_{xy}=2t^{*2}/Ud$ and $J_{z}=2t^{*2}/Ud+V^{*}/d$.
Therefore, an infinitesimally weak $V^{*}>0$ is sufficient to break a pure SC in $U\rightarrow \infty$.

\begin{figure}[bht]
\begin{center}
\epsfig{file=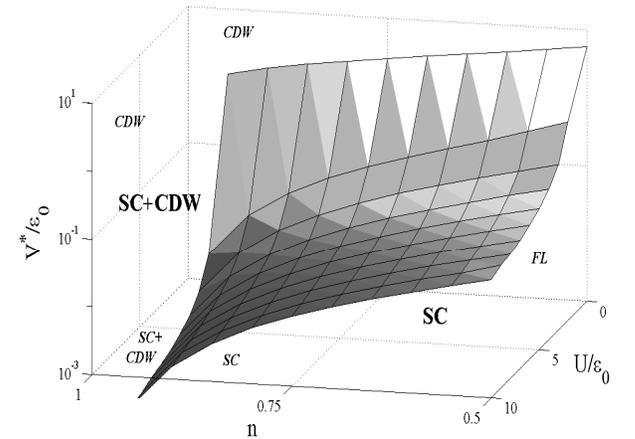,width=8cm,height=6cm}
\caption{
Phase diagram of the extended attractive Hubbard model in $d=\infty$.
The curve is the phase boundary of a pure SC and a coexistence phase (SC+CDW).
Italic indicates a phase on a plain.
} 
\label{Phase}
\end{center}
\end{figure}

The phase diagram of the infinite dimensional extended attractive Hubbard model (\ref{tUV}) is shown in Fig. \ref{Phase}.
At $V^{*}=0$~\cite{Micnas90} or $U=0$~\cite{Uhrig93}, our result is the same as the MF results.
The rest of the phase boundaries are different from those of simple MF theories.
The fact that we have found a quantum phase transition from a pure SC, proves the existence of a {\it discontinuous} evolution.
The BCS theory is one of the most successful theories in physics, so this is a significant phenomenon.
The quantum phase transition is an attractive version of the Mott transition.

In summary, we have made a variational analysis on an extended attractive Hubbard model in $d=\infty$.
We have found the importance of electronic correlations through evaluations of superconducting condensation energies.
As far as a pure superconducting phase is concerned, an evolution with increasing attraction turns out to be continuous in a dilute system.
However, such a smooth evolution scenario breaks down near half filling, where charge density waves take place.
We have found a quantum phase transition between a pure superconducting phase and a coexistence phase of superconductivity and charge density waves induced by varying interactions and/or filling.
This phenomenon can be viewed as a Mott transition in an attractive system.

We appreciate enlightening discussions with Prof. I. Terasaki.
This work is supported by Waseda University Grant for Special Research Project (98A-855, 99A-564).

\end{multicols}

\end{document}